\documentclass[11pt]{article}
\usepackage[utf8]{inputenc}

\usepackage{url} 
\usepackage{lineno}
\usepackage{graphicx}
\usepackage{hyperref}
\usepackage[affil-it]{authblk}
\usepackage{soul}
\usepackage{amsfonts}  
\usepackage{booktabs}  
\usepackage[margin=1in]{geometry}
\usepackage[table]{xcolor}
\usepackage{amsmath}

\usepackage{natbib}

\begin{document}
\title{\hrule \vspace{15pt}\textbf{BharatBench: Dataset for data-driven weather forecasting over India} \vspace{15pt} \hrule}

\author[1]{Animesh Choudhury}
\author[1,*]{Jagabandhu Panda}
\author[1]{Asmita Mukherjee}

\affil[1]{Department of Earth and Atmospheric Sciences, National Institute of Technology, Rourkela, India}
\affil[*]{Corresponding author: jagabandhu@gmail.com}
\date{}

\maketitle

\begin{abstract}
Advanced weather and climate models use numerical techniques on grided meshes to simulate atmospheric and ocean dynamics, which are computationally expensive. Data-driven approaches are gaining popularity in weather and climate modeling, with a broad scope of applications. Although Machine Learning (ML) has been employed in this domain, significant progress has occurred in the past decade, leading to ML applications that are now competitive with traditional numerical methods. This study presents a user-friendly dataset for data-driven medium-range weather forecasting focused on India. The dataset is derived from IMDAA reanalysis datasets and optimized for ML applications. The study provides clear evaluation metrics and a few baseline scores from simple linear regression techniques and deep learning models. The dataset can be found at \url{https://www.kaggle.com/datasets/maslab/bharatbench}, while the codes are available at \url{https://github.com/MASLABnitrkl/BharatBench}. We hope this dataset will boost data-driven weather forecasting over India. We also address limitations in the current evaluation process and future challenges in data-driven weather forecasting.
\end{abstract}

\begin{center}
    \line(1,0){250}
\end{center}

\section{Introduction}

Weather forecasting has gained considerable attention across various disciplines due to its profound implications on daily life, including agriculture, commerce, emergency management,  etc. Most current weather prediction systems are based on physical principles that solve governing equations on a meshed grid \citep{de2023}. Although this approach has delivered great success, the computational cost of running these models is substantial, especially for ensemble forecasts \citep{ben2024, Hobeichi2023}. Given these challenges and the escalating interest in machine learning (ML), there has been a surge in adopting data-driven approaches to enhance and expedite numerical weather prediction (NWP). Deep learning (DL), a subset of ML characterized by multi-layered artificial neural networks (ANNs), has emerged as a potent tool across a spectrum of weather and climate applications\citep{de2023, mcgovern2023}. The efficacy of DL largely relies on the capacity of neural networks to identify patterns within intricate, high-dimensional data spaces. The pursuit of data-driven forecasting, extending from language translation to audio signal processing and even numerical simulations, underscores an evolving area of research \citep{mcgovern2023, slater2023}. Within this expansive landscape, weather forecasting emerges as a particularly formidable challenge. Data-driven models could learn more efficient ways to predict the weather, making forecasting more reliable \citep{dueben2018}. They could also help create large ensembles to better predict extreme events. Data-driven models may even outperform traditional physical models in challenging areas. Additionally, new tools in Explainable artificial intelligence (XAI) methods could provide better insight \citep{bommer2023}. Finally, there's a scientific question about how well data-driven models can understand atmospheric dynamics.

Researchers have recently explored the potential of data-driven approaches for mesoscale NWP, and the initial results are encouraging. The integration of these approaches has shown great promise in improving forecasting capabilities by improving accuracy, spatial and temporal resolution, and offering tailored prediction \citep{de2023}. \cite{dueben2018} developed a toy model based on a Neural Network (NN) to understand the key challenges and design considerations for achieving optimal results. \cite{scher2018} and \cite{scher2019} employed Convolutional Neural Networks (CNNs) with an encoder-decoder setup to predict 3D model fields. \cite{scher2018} trained separate networks for various lead times up to 14 days, while in the study of \cite{scher2018}, training was focused on 1-day forecasts, which were then extended iteratively. \cite{weyn2019} trained deep CNNs to predict reanalysis-derived 500 hPa and 700-300 hPa geopotential thickness data at 6-hour intervals. The NN design resembled the encoder-decoder convolutional networks employed in previous studies by \cite{scher2018} and \cite{scher2019}. The studies also conducted experiments by introducing a convolutional long short-term memory (LSTM) hidden layer into the network architecture.

These promising attempts demanded for the preparation of the benchmark datasets such as ImageNet \citep{deng2009}, GLUE \citep{wang2018},  MNIST \citep{lecun1998}, SuperBench \citep{ren2023}, and many more (\url{https://paperswithcode.com/datasets}). \cite{Rasp2020} prepared a benchmark dataset (WeatherBench) using the ERA5 archive to facilitate the application of ML/DL, along with a robust evaluation matrix and some baseline models. Since its inception, several studies have explored its potential and developed different data-driven approaches. \cite{weyn2019} utilized a cubed sphere projection and a UNet architecture for iterative atmospheric state prediction at a 2-degree resolution. \cite{rasp2021} employed a deep Resnet architecture to forecast fields up to 5 days ahead with a resolution of 5.625 degrees. \cite{clare2021} also used a Resnet but integrated a probabilistic output layer. Additionally, a probabilistic extension to WeatherBench, including various machine learning (ML) baselines, was presented by \cite{garg2022}.

Advancements in data-driven weather modeling have gained momentum since the 2020s. \cite{keisler2022} employed a graph neural network (GNN) with a 6-hour time step, 1◦ resolution, and 13 vertical levels, showing comparable skills to operational NWP models. \cite{pathak2022} introduced FourCastNet, a modified vision transformer, for high-resolution prediction. Subsequently, an enhanced version incorporating spherical Fourier neural operators was recently introduced by \cite{bonev2023}. Pangu-Weather \citep{bi2023}, using a unique variation of vision transformers at 0.25◦ spatial resolution, found better performance than the operational Integrated Forecasting System (IFS). Following this, GraphCast \citep{lam2023}, based upon \cite{keisler2022}, by adopting a GNN to a 0.25◦ horizontal resolution, outperforms IFS on many occasions, especially in indicating extreme weather. In 2023, a new type of vision transformer called FengWu \citep{chen2023a} was introduced, showing excellent performance in long-term forecasts. The FuXi model \citep{chen2023b} matches the accuracy of the IFS ensemble mean for up to 15 days. SwinRDM \citep{chen2023c} combines two types of networks to make detailed predictions at a resolution of 0.25◦. ClimaX \citep{nguyen2023a} and Stormer \citep{nguyen2023b}(Nguyen et al., 2023) are models that perform well at lower resolutions. Finally, NeuralGCM \citep{hoyer2023} is the first model to blend ML and physics, achieving top-notch results, even in an ensemble format.

The WeatherBench dataset, derived from the global ERA5 dataset, offers broad coverage of meteorological information but may lack regional specificity and resolution. India has diverse landscapes, from deserts to high-altitude mountains and fertile plains to plateaus, resulting in various microclimates. This diversity influences its weather patterns, causing unique conditions in different areas. However, this diversity also exposes India to various extreme weather events like heat waves, cyclones, droughts, thunderstorms, and more. Recently, the Ministry of Earth Sciences (MoES) of India created a special center focused on applying AI and ML to improve weather, climate, and ocean prediction skills. Additionally, a computing setup and virtual workspace have been established at the Indian Meteorological Department (IMD) for training and deploying data-driven models utilizing graphical processor-based servers. MoES envisions a future where weather and climate forecasts blend AI/ML models with traditional numerical prediction methods. Institutes under MoES are consistently urged to embrace AI and ML advancements in Earth Sciences. To support this, MoES is dedicated to improving the High-Performance Computing (HPC) infrastructure. In this study, we propose the creation of "BharatBench," a specialized dataset tailored for ML/DL applications focused specifically on the Indian region. BharatBench is curated using the IMDAA (Indian Meteorological Department Analysis and Assimilation) reanalysis dataset, ensuring greater regional detail and accuracy \citep{rani2021}. BharatBench includes baseline models and evaluation metrics for comprehensive analysis and model comparison. The article begins with an overview of prior research in the field (section 1), followed by a discussion on the preparation of the dataset (section 2) and the assessment criteria utilized (Section 3). Subsequently, various baseline models are described (Section 4). The paper also identifies potential avenues for future investigation (Section 5) and concludes by offering a comprehensive perspective (Section 6).

\section{Dataset}
\label{sec:data}
The IMDAA reanalysis dataset \citep{rani2021} offers detailed meteorological information, available at a spatial grid of 0.12 degrees with 24 vertical levels from 1979 onwards (\url{https://rds.ncmrwf.gov.in/datasets}). Reanalysis datasets combine simulation results with observations, providing a reliable representation of atmospheric conditions over time. These datasets are characterized by long-term continuity and high resolution, offering valuable insights into meteorological variables. Since the volume of the raw data is quite large, the proposed dataset was prepared for the period 1990 to 2020 and clipped for 5°N to 40°N and 65°E to 100°E. The data was regrided to a lower resolution of 1.08° (32 × 32 grid points), which aided in handling memory and processing constraints. The dataset encompasses surface variables, atmospheric variables across 13 vertical pressure levels (50, 100, 150, 200, 250, 300, 400, 500, 600, 750, 850, 925, and 1000 hPa), and constant fields listed in Table ~\ref{tab:Table1}. Using vertical pressure coordinates enhances simplicity and facilitates mass conservation in the dataset, aligning with established meteorological principles. Each surface variable and constant contributes around 182MB to the dataset, while each atmospheric variable add another 2.24GB. You can access the dataset in NetCDF format at \url{https://www.kaggle.com/datasets/maslab/bharatbench}.

\begin{figure}
\includegraphics[width=\textwidth]{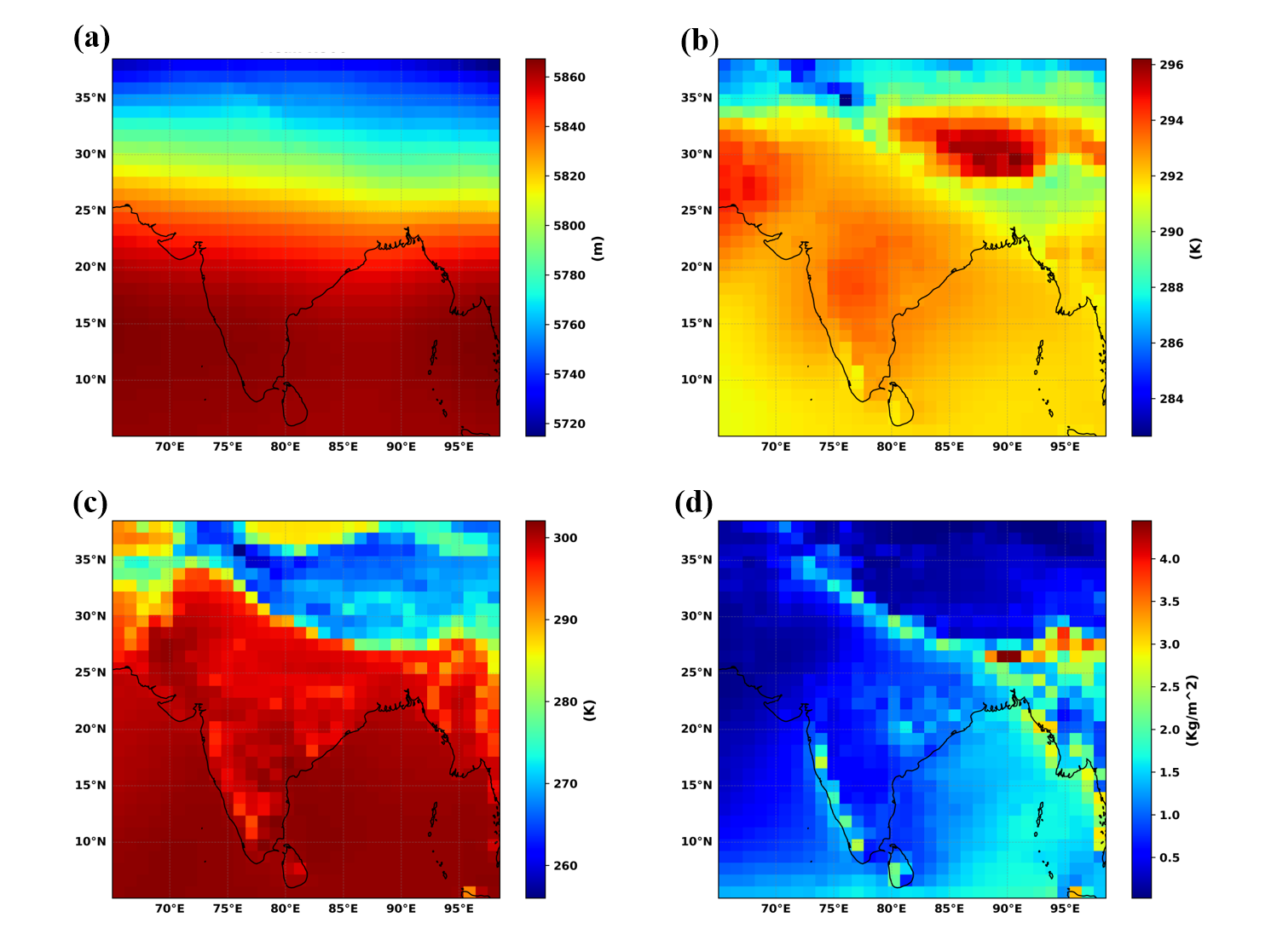}
\caption{Spatial distribution of average (a) H500, (b) T850, (c) T2m, and (d) TP6h.}
\label{fig:Figure1}
\end{figure}

For the preliminary analysis, we considered four variables: geopotential height at 500 hPa (H500), temperature at 850 hPa (T850), temperature at 2m (T2m), and six hourly accumulated precipitation (TP6h). The average values of these four variables are shown in Figure~\ref{fig:Figure1}. H500 provides an approximate altitude at which the atmospheric pressure drops to 500 hPa, typically around 5.5 km above sea level. This level, known as the steering level, indicates the general direction of weather systems as they align with winds at the 500 hPa level. The 850 hPa level is around 1.5 km above the sea level and slightly above the boundary layer. In general, the variation in temperature during day and night is insignificant at this level. So, air temperature at this level can be used to differentiate between warm and cold fronts. The 2-meter surface temperature field gives forecasters an overview of current and future temperatures at the Earth's surface. Predicting the amount of precipitation in a specific area is crucial for agricultural planning, flood control, and understanding the water cycle's environmental impact, etc. Precise measurement of precipitation aids scientists in tracking trends and making climate predictions. Moreover, accurate precipitation data is vital for public safety as it contributes to storm warnings and helps individuals make informed decisions about evacuating areas prone to hazardous weather conditions. Concerning data complexity, H500 generally exhibits stable and consistent trends over short time intervals, reflecting large-scale atmospheric circulation patterns, T850 shows moderate variability in trends, influenced by air mass movements, frontal boundaries, and synoptic weather systems, T2m exhibits highly dynamic and fluctuating trends, following diurnal and seasonal variations along with local weather conditions, and TP6h displays irregular and sporadic trends, characterized by peaks during precipitation events and relatively flat periods in dry conditions.

\begin{table*}[]
\small
\centering
\begin{tabular}{cccc}
\toprule
\textbf{Variable}        & \textbf{Short name} & \textbf{Unit} & \textbf{Levels} \\ \midrule
\multicolumn{4}{c}{\textit{Surface variables}} \\                                                                   
2m Temperature                & TMP\_2m                & K                         & 1        \\
10m u component of wind       & UGRD\_10m              & m/s                       & 1        \\
10m v component of wind       & VGRD\_10m              & m/s                       & 1         \\
Mean sea level pressure       & PRMSL\_msl             & Pa                        & 1         \\     
Total Precipitation           & APCP\_sfc              & Kg/m$^2$                    & 1         \\ 

\midrule
\multicolumn{4}{c}{\textit{Atmospheric variables}}                                                                          \\
Temperature                & TMP\_prl                & K                          & 13                          \\
Relative humidity          & RH\_prl                 & \%                         & 13                          \\
Geopotential height        & HGT\_prl                & m                          & 13                          \\
U component of wind        & UGRD\_prl               & m/s                        & 13                         \\
V component of wind        & VGRD\_prl               & m/s                        & 13                         \\
\midrule
\multicolumn{4}{c}{\textit{Constants}}                                                                          \\
Land sea mask            & Land\_sfc                 & [0/1]                          & 1                       \\
Terrain Height           & MTERH\_sfc                & m                              & 1                        \\

\bottomrule
\end{tabular}
\caption{List of selected variables contained in this dataset.}
\label{tab:Table1}

\end{table*}

\section{Methodology}
\label{sec:Method}
In this study, the whole dataset was divided into training (1990 to 2017), validation (2018), and testing (2019 to 2020) datasets. The training set was used to train and fit the model, while the validation set was used to assess the model’s performance during the training process. However, there is a risk of unintentionally overfitting the validation set if hyperparameters are continuously tuned. The test set was kept aside to check the model’s final performance. Since meteorological variables exhibit high temporal correlation, choosing a longer continuous data segment for validation is preferable over a random split.
To compare the performance of the different forecasting models, it is essential to have a robust, acceptable, and computationally easy evaluation matrix. Here, root mean square error (RMSE), mean absolute error (MAE), and anomaly correlation coefficient (ACC) were estimated to check the accuracy of the models. RMSE and MAE both measured the distance between the actual and predicted values, but RMSE has more sensitivity towards outliers compared to MAE. However, when outliers are exponentially rare, like in a bell-shaped curve, RMSE is preferred. The RMSE and MAE were computed as:
\begin{equation*}
\mathrm{RMSE}=\frac{1}{N_{\mathrm{pred\ }}}\sum_{i}^{N_{\mathrm{pred\ }}}\sqrt{\frac{1}{N_{\mathrm{lat\ }}N_{\mathrm{lon\ }}}\sum_{j}^{N_{\mathrm{lat\ }}}\sum_{k}^{N_{\mathrm{lon\ }}}\left(f_{i,j,k}-t_{i,j,k}\right)^2}
\end{equation*}
\begin{equation*}
\mathrm{MAE\ }=\frac{1}{N_{\mathrm{pred\ }}}\sum_{i}^{N_{\mathrm{pred\ }}}\frac{1}{N_{\mathrm{lat\ }}N_{\mathrm{lon\ }}}\sum_{j}^{N_{\mathrm{lat\ }}}\sum_{k}^{N_{\mathrm{lon\ }}}\left|f_{i,j,k}-t_{i,j,k}\right|
\end{equation*}
where ` f ' is the model prediction, and `t' is the actual value.

ACC, one of the most widely used metrics for verifying spatial fields, measures the spatial correlation between anomalies of forecasts and those of verifying values with climatological values. ACC indicates how well the forecasted anomalies match the observed anomalies, reflecting the accuracy of the forecast model. ACC values range from +1 to -1, where values near +1 signify good agreement, indicating valuable forecast anomalies; values around 0.5 suggest forecast errors similar to those based on a climatological mean; values near 0 indicate poor agreement and negligible forecast value, and values approaching -1 indicate anti-phase agreement and unreliable forecasts. The ACC was defined as:
\begin{equation*}
\mathrm{ACC}=\frac{\sum_{i,j,k}{f_{i,j,k}^\prime t_{i,j,k}^\prime}}{\sqrt{\sum_{i,j,k} f_{i,j,k}^{\prime2\sum_{i,j,k} t_{i,j,k}^{\prime2}}}}
\end{equation*}
where the prime $^\prime$  denotes the difference to the climatology.

\section{Results and discussion }
\label{sec:Results}
Baseline models are essential in data-driven research as they provide a reference point for evaluating more complex models. This study provides a few baseline models along with their evaluation scores. The baseline models tried to predict a variable (H500, T850, T2m, TP6h) three to five days ahead and were evaluated based on RMSE, MAE, and ACC. These baselines serve as a starting point to assess the effectiveness of new algorithms. The spatial distribution of these variables (Figure~\ref{fig:Figure1}) had some notable patterns. H500 varied more with latitude than longitude, and also showed higher variation in areas with lower mean H500 values. T2m had a stronger correlation with land cover than T850 and exhibited greater daily fluctuations. TP6h had very high standard deviation values comparable to its mean value.

\begin{figure} 
\includegraphics[width=\textwidth]{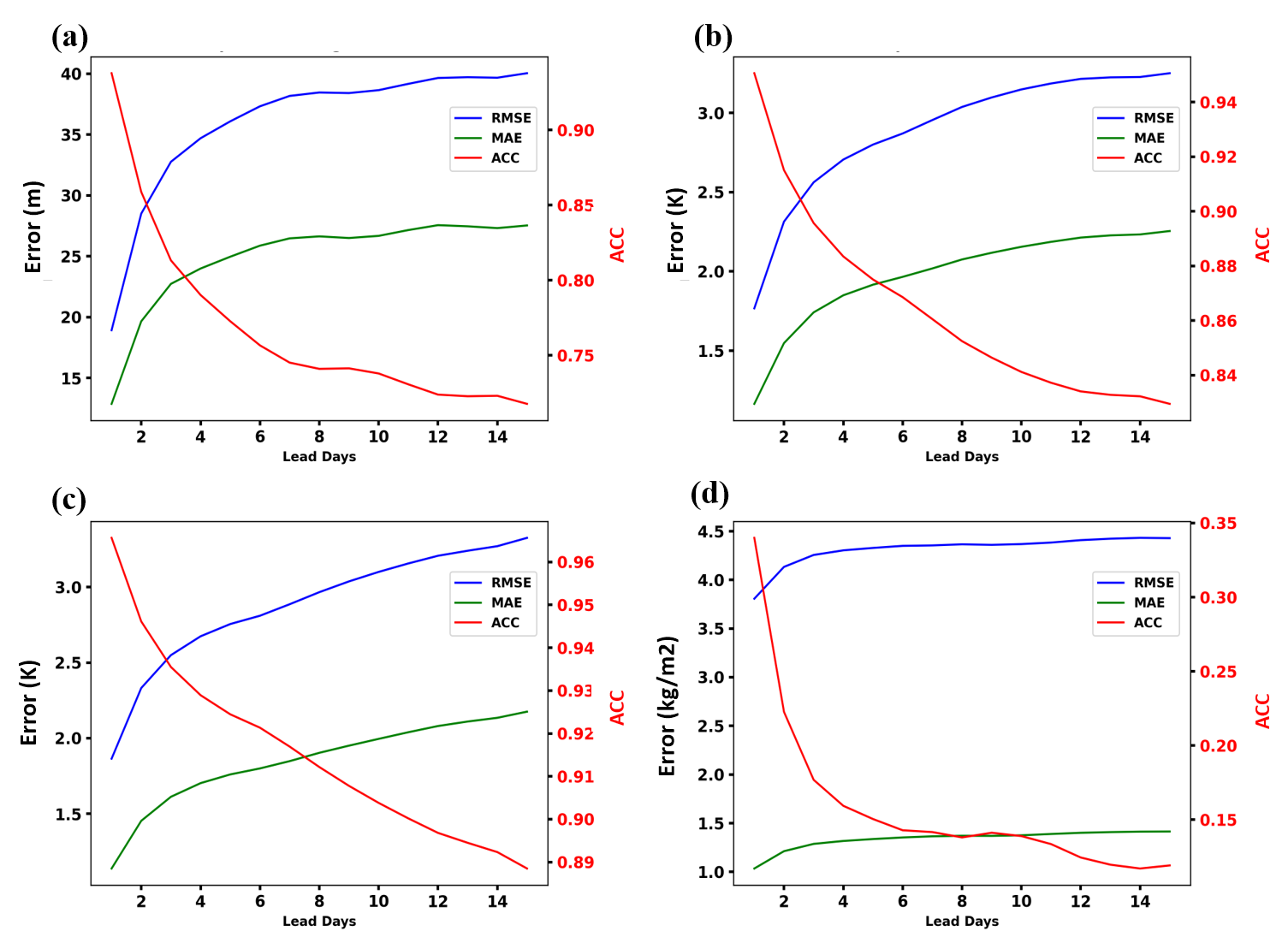}
\caption{RMSE MAE and ACC  for persistence forecasting of (a) H500, (b) T850, (c) T2m, and (d) TP6h at different lead times. The blue and green lines represent RMSE and MAE, respectively, while the red line depicts the Accuracy (ACC).
}
\label{fig:Figure2}
\end{figure}

We started with a persistence forecast model, a simple but important baseline model used in weather forecasting. It predicts future values of a variable based on its current or most recent value, assuming no change or persistence in the pattern. We investigated the persistence forecasting up to 15 days ahead and observed that errors (RMSE, MAE) in the prediction became larger as we forecast further into the future (Figure\ref{fig:Figure2} ). The errors increased rapidly up to 7 days, after which the rate of increase slowed down. Lower error values signify better model accuracy, while a decreased ACC suggests poor performance.

Climatological forecasting is another simple forecasting model that involves computing climatologies from historical data to establish baseline expectations for future weather conditions. The climatology was computed in two ways using the entire training data set spanning from 1990 to 2017. One method calculated the average over the entire training period, while the other computes averages for each week in a year, reflecting seasonal changes. The results showed that weekly climatology performed similarly to the climatology and closely resembled the persistence forecast two to three days ahead (Table ~\ref{tab:Table2}).

\begin{table*}[]
\centering
\begin{tabular}
{
>{\columncolor[HTML]{FFFFFF}}c 
>{\columncolor[HTML]{FFFFFF}}c 
>{\columncolor[HTML]{FFFFFF}}c 
>{\columncolor[HTML]{FFFFFF}}c 
>{\columncolor[HTML]{FFFFFF}}c 
>{\columncolor[HTML]{FFFFFF}}c 
}
\toprule
\textbf{Climatology}        & \textbf{\begin{tabular}[c]{@{}c@{}}Evaluation \\ Matrix\end{tabular}} & \textbf{H500} & \textbf{T850} & \textbf{T2m} & \textbf{TP6h} \\
\midrule
                            & RMSE                                                                  & 28.474812     & 2.517884      & 4.135361     & 3.177998      \\
\textbf{}                   & MAE                                                                   & 21.135204     & 1.781713      & 2.827197     & 1.116252      \\
                            & ACC                                                                   & 0.8539703     & 0.894133      & 0.814178     & 0.286481      \\
                            &                                                                       &               &               &              &               \\
\toprule                            
\textbf{Weekly Climatology} & \textbf{\begin{tabular}[c]{@{}c@{}}Evaluation \\ Matrix\end{tabular}}                                                                      & \textbf{H500} & \textbf{T850} & \textbf{T2m} & \textbf{TP6h} \\ 
\midrule
                            & RMSE                                                                  & 28.775707     & 2.537314      & 4.150869     & 3.166719      \\
                            & MAE                                                                   & 21.278952     & 1.794695      & 2.835194     & 1.112704      \\
                            & ACC                                                                   & 0.8503808     & 0.892405      & 0.81263      & 0.294683     

\end{tabular}

\caption{Error matrices for climatology and weekly climatology of the selected variable.}
\label{tab:Table2}
\end{table*}

For a forecast system to be deemed valuable, it must outperform climatology, weekly climatology, and persistence forecasts, emphasizing the need for advanced modeling techniques. A linear regression model predicted the four selected variables 3 and 5 days ahead. The spatial fields were flattened from 32x32 to 1024. The findings showed that the linear regression model had notable success in predicting three days in advance, and even its predictions at five days ahead were comparable to those of the climatological forecast (Table~\ref{tab:Table3}).

\begin{table*}[]
\centering
\begin{tabular}{
>{\columncolor[HTML]{FFFFFF}}c 
>{\columncolor[HTML]{FFFFFF}}c 
>{\columncolor[HTML]{FFFFFF}}c 
>{\columncolor[HTML]{FFFFFF}}c 
>{\columncolor[HTML]{FFFFFF}}c 
}
\toprule
3days         & \textbf{H500} & \textbf{T850} & \textbf{T2m} & \textbf{TP6h} \\
\midrule
\textbf{RMSE} & 26.61557      & 2.005987      & 3.214307     & 2.10935       \\
\textbf{MAE}  & 18.743786     & 1.396874      & 1.217106     & 1.363918      \\
\textbf{ACC}  & 0.8685174     & 0.933918      & 0.308317     & 0.95493       \\
\toprule
5days         & \textbf{H500} & \textbf{T850} & \textbf{T2m} & \textbf{TP6h} \\
\midrule
\textbf{RMSE} & 29.582294     & 2.202205      & 3.265312     & 2.2735        \\
\textbf{MAE}  & 20.688204     & 1.539523      & 1.245188     & 1.47812       \\
\textbf{ACC}  & 0.83359593    & 0.919551      & 0.272551     & 0.947356     \\
\end{tabular}
\caption{Error matrices for Linear Regression of the selected variable for 3days and 5days forecasts.}
\label{tab:Table3}
\end{table*}

\begin{figure}[h]
\includegraphics[width=\textwidth]{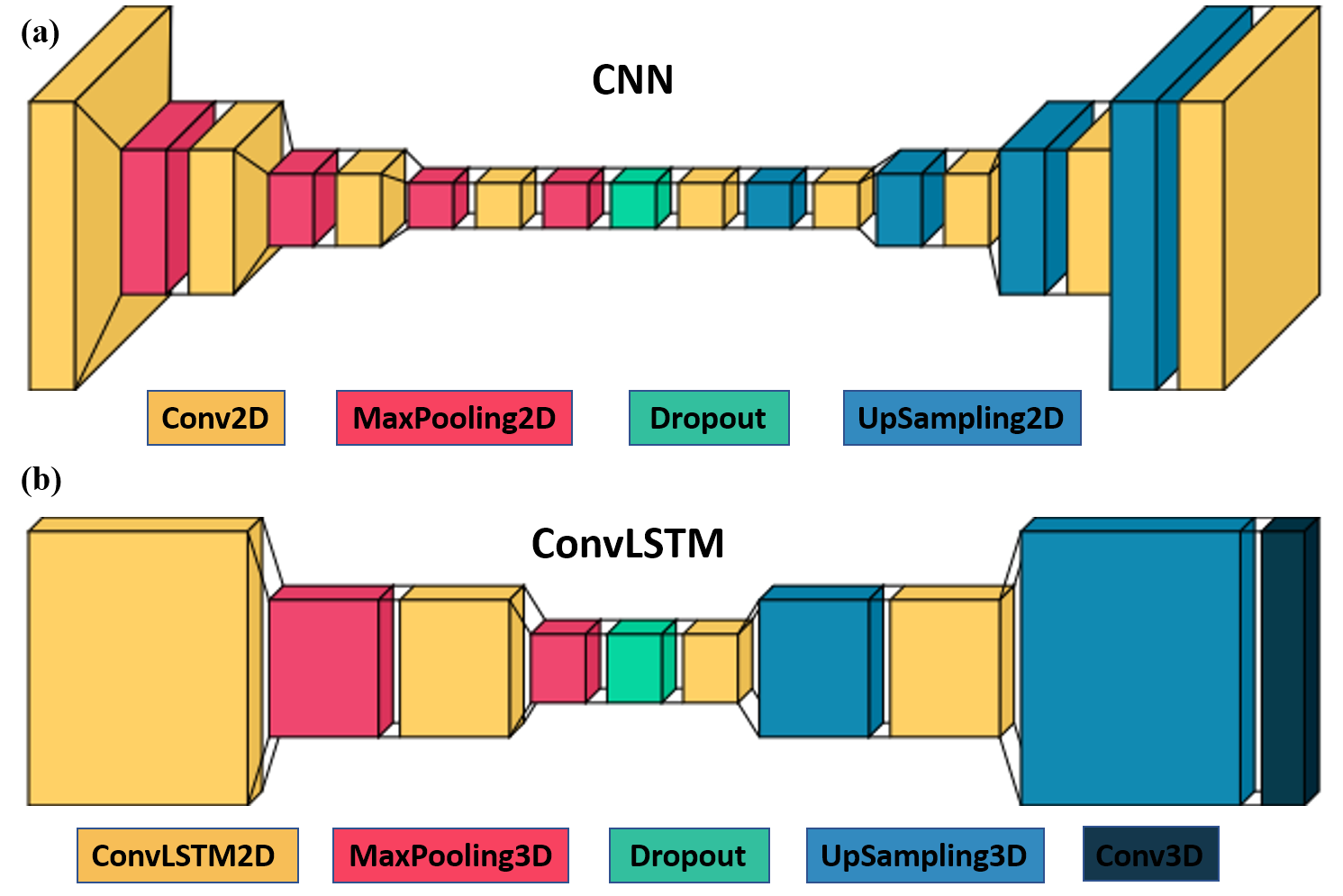}
\caption{Graphical representation of the (a) CNN and (b) ConvLSTM baseline architecture.}
\label{fig:Figure3}
\end{figure}

CNNs are a natural choice for spatial data due to their ability to effectively capture patterns and features regardless of their location or orientation within the data. Additionally, CNNs automatically learn a hierarchical representation of features, starting from simple features at lower layers to complex information from spatial data. We trained a CNN with 17 layers, including eight convolutional layers, four max-pooling layers, four upsampling layers, and a dropout layer. The dropout layer was used to avoid the model being overfitted. Each convolutional layer used 32 filters with a kernel size of 5 and a “swish” activation function. The model was trained separately for H500 and T850 using the Adam optimizer and a mean squared error loss function. The CNN forecasts for 3- and 5-days outperformed climatology and weekly climatology but struggled to surpass linear regression (Table ~\ref{tab:Table4}). The H500 forecasts were slightly better than linear regression after extensive training, while T850 forecasts fell slightly short compared to linear regression. ConvLSTM combines the spatial capabilities of CNNs with the temporal sequencing of LSTMs (Long Short Term Memory), making it suitable for analyzing sequences of spatial data over time. We trained a ConvLSTM model consisting of 4 ConvLSTM2D layers, two upsampling and max-pooling layers each, and one dropout layer. This model's total number of trainable parameters was 186,721, and all other parameters remained similar to the CNN model. The model's performance was close to the linear regression and CNN but not better (Table ~\ref{tab:Table4}). The architecture of the CNN and ConvLSTM are show in the Figure ~\ref{fig:Figure3}. 

During neural network training, we first trained the model to predict the variable three days ahead and then extended the training to predict five days ahead. Interestingly, the training time for the latter was notably shorter, indicating the effectiveness of transfer learning. Additionally, we observed the importance of using an adaptive learning rate. Initially, training with an appropriate learning rate reduced the validation loss. However, as training progressed, the validation loss would eventually diverge from its minimum. To enhance model performance, it was necessary to continue training with a lower learning rate at this point. Adding more convolutional layers increases the number of trainable parameters, making the model take longer to train with the same computational power. However, increasing the layers could lead to improved results. The variables we compared in the baseline models (H500, T850) had a strong linear relationship. This might explain why the neural network struggled to improve performance. It was hard to find a non-linear function with a neural network that could represent the time evolution of the selected variables. The address for the code repository is \url{https://github.com/MASLABnitrkl/BharatBench}, which allows for result reproduction and an easy start for users.
\begin{table*}[]
\centering
\begin{tabular}{
>{\columncolor[HTML]{FFFFFF}}c 
>{\columncolor[HTML]{FFFFFF}}c 
>{\columncolor[HTML]{FFFFFF}}c 
>{\columncolor[HTML]{FFFFFF}}c 
>{\columncolor[HTML]{FFFFFF}}c 
>{\columncolor[HTML]{FFFFFF}}c
}
\toprule
\textbf{Models} & \textbf{\begin{tabular}[c]{@{}c@{}}Evaluation \\ Matrix\end{tabular}} & \multicolumn{2}{c}{\textbf{3 days}} & \multicolumn{2}{c}{\textbf{5days}} \\
\midrule
                &                                                                       & \textbf{H500}                & \textbf{T850}                & \textbf{H500}                & \textbf{T850}               \\
\midrule                
CNN    & RMSE                                                                  & 26.27218                     & 2.053407                     & 28.805109                    & 2.256589                    \\
                & MAE                                                                   & 18.427433                    & 1.433121                     & 20.107533                    & 1.577852                    \\
                & ACC                                                                   & 0.8724568                    & 0.930596                     & 0.8438257                    & 0.915599                    \\
\midrule                
ConvLSTM        & RMSE                                                                  & 26.87115                     & 2.210813                     & 29.592497                    & 2.381422                    \\
                & MAE                                                                   & 18.805605                    & 1.535963                     & 20.747108                    & 1.657516                    \\
                & ACC                                                                   & 0.8660521                    & 0.919435                     & 0.8346201                    & 0.905175      \\             
\end{tabular}
\caption{Error matrices for CNN and ConvLSTM of the selected variable for 3days and 5days forecasts .}
\label{tab:Table4}
\end{table*}

\section{Limitations and Future scope}
\label{sec:Limit}
The task of predicting the state of atmospheric variables based on the current condition is similar to image-to-image translation in ML. Just as image-to-image translation models learn to understand and transform visual patterns, state-to-state weather prediction models learn to capture and predict patterns and changes in atmospheric variables over time.  However, forecasting weather differs from typical image-to-image applications in several ways, leading to unique challenges.

Weather forecasting involves predicting how atmospheric conditions will evolve over time. This temporal aspect introduces a dynamic element that is not present in static image-to-image translation tasks. Models must capture not just spatial patterns but also temporal changes such as diurnal cycles, movement of weather fronts, and seasonal variations. Weather data is three-dimensional, but traditional CNNs treat different levels as separate channels, which may not capture vertical dynamics effectively. Using a 3D CNN could be problematic due to changing atmospheric dynamics and grid spacings in the vertical direction, violating the assumption of translation invariance. Moreover, dynamics vary with latitude due to stretched grid cells and the Coriolis effect. Solutions could include the use of spherical convolutions or incorporating latitude information into the network to address these challenges. 

Another constraint is the limited training data that raises concerns about overfitting complex networks. Transfer learning, where networks are pretrained on similar tasks or datasets like climate simulations, then fine-tuned, can help address this. Data augmentation, common in computer vision but tricky for physical fields, can also prevent overfitting. Techniques like random rotations may not work well due to the distinct x and y directions in physical data. Ensemble analyses and forecasts can add diversity to the training data, helping to tackle this challenge. On the other hand, storing and loading this huge amount of data efficiently can be a bottleneck, especially for high-resolution models or multiple variables.

The dataset does not address probabilistic forecasting, crucial for understanding forecast uncertainty due to atmospheric chaos. In traditional models, this is achieved through ensembles of slightly varied initial conditions and model physics. Extending data-driven models to produce probabilistic forecasts involves computing probabilistic scores, which is an intriguing research area. Researchers are encouraged to explore this aspect in the dataset to enhance forecast accuracy. Also, weather events like heat waves or cold waves are rare but have significant societal impacts. These events are not well captured by regular metrics like RMSE. While evaluating extreme situations separately makes sense, there is no standard metric for defining extremes. Users are encouraged to define and evaluate extremes based on their specific needs. One of the many ways to improve data-driven weather forecasting is to increase the training data and explore more complex network architecture. The dataset provides diverse data variables that have not been fully explored, and it is hoped that future studies will reveal the most effective variable combinations.

\section{Conclusion}
\label{sec:conclude}
The advancements in ML/DL techniques have accelerated their adoption across various domains, including weather forecasting, marking significant progress in this field. As a developing nation, India needs an accurate yet computationally efficient weather prediction system. A parallel data-driven approach could serve as a viable alternative either entirely or in part to the country's current weather forecasting system. This study introduces a readily available dataset tailored for ML/DL applications in weather and climate studies focused on India. Leveraging a data-driven approach holds promise for addressing limitations associated with conventional NWP methods. We anticipate that this dataset will serve as a cornerstone for speedy research in this field. 

Weather forecasting impacts many dimensions of society. Current weather prediction systems require huge computational resources and systematic maintenance and upgrades. The data-driven approach has emerged as an economical solution, although there is doubt about the capability of a completely data-driven method to accurately represent atmospheric conditions. To make data-driven approaches reliable, it is imperative that they not only exhibit exceptional predictive performance, often measured by high accuracy, but also furnish interpretable insights that align with established scientific theory \citep{mcgovern2019, thiebes2021}.

This project is hosted on Github, serving as a central hub for collaboration, code sharing, and discussions through Github issues, fostering a collaborative and transparent research environment. We have established a clear measure of success for our data-driven, medium-range forecast model by using RMSE scores for 500 hPa geopotential height and 850 hPa temperature fields. This measure simplifies the evaluation process, providing a straightforward assessment of the model's performance. Furthermore, the code repository offers a quick-start Jupyter notebook that guides users through data handling, model training, and result assessment, making it easy for researchers interested in the work to get started.

\section*{Acknowledgment}
Authors gratefully acknowledge NCMRWF, Ministry of Earth Sciences, Government of India, for IMDAA reanalysis. IMDAA reanalysis was produced under the collaboration between the UK Met Office, NCMRWF, and IMD with financial support from the Ministry of Earth Sciences under the National Monsoon Mission program.

\newpage
\small

\bibliographystyle{apalike}



\end{document}